\pdfoutput=1
\documentclass[5p,times]{elsarticle}

\def\BibTeX{{\rm B\kern-.05em{\sc i\kern-.025em b}\kern-.08em
    T\kern-.1667em\lower.7ex\hbox{E}\kern-.125emX}}
\usepackage{amsmath,dsfont, bm}
\usepackage{amssymb,amsthm}
\usepackage{epsfig,verbatim}
\usepackage{setspace}
\usepackage{color}
\usepackage{epstopdf}
\usepackage{graphics}
\usepackage{accents}
\usepackage{acronym}
\usepackage[bookmarks,breaklinks, colorlinks,pdfencoding=auto, psdextra]{hyperref}
\usepackage{soul}
\usepackage{enumitem}
\usepackage{mathrsfs}
\usepackage{bookmark}
\usepackage{algpseudocode}
\usepackage{algorithm}
\usepackage{float}
\usepackage{booktabs}
\usepackage{stackengine}
\usepackage{subcaption}
\usepackage{scalefnt}
\usepackage{xcolor}
\usepackage{url}
\usepackage{hyperref}
\usepackage{multirow}
\usepackage{pgfplots}
\usepgfplotslibrary{groupplots}
\usepgfplotslibrary{statistics}
\pgfplotsset{compat=1.18}
\usepackage{array} 
\usepgfplotslibrary{fillbetween}

\usepackage{tikz}
\usetikzlibrary{shapes.geometric, arrows.meta, positioning, calc}

\acrodef{sh}[SH]{Spherical Harmonics}
\acrodef{hrtf}[HRTF]{Head Related Transfer Function}
\acrodef{g}[G]{Global}
\acrodef{t}[T]{Time}
\acrodef{tf}[TF]{Time Frequency}
\acrodef{tsh}[TSH]{Time Spherical Harmonics}
\acrodef{tfsh}[TFSH]{Time Frequency Spherical Harmonics}
\acrodef{ar}[AR]{Augmented Reality}
\acrodef{vr}[VR]{Virtual Reality}
\acrodef{stft}[STFT]{Short Time Fourier Transform}
\acrodef{sht}[SHT]{Spherical Harmonic Transform}
\acrodef{isht}[ISHT]{Inverse Spherical Harmonic Transform}
\acrodef{foa}[FOA]{First Order Ambisonics}
\acrodef{hoa}[HOA]{Higher Order Ambisonics}
\acrodef{drr}[DRR]{Direct-to-Reverberant Ratio}
\acrodef{iacc}[IACC]{Interaural Cross-Correlation}
\acrodef{sdr}[SI-SDR]{Scale-Invariant Signal-to-Distortion Ratio}
\acrodef{ift}[IFT]{Inverse Fourier Transform}
\acrodef{dirac}[DirAC]{Directional Audio Coding}
\acrodef{compass}[COMPASS]{Coding and Multidirectional Parameterization of Ambisonic Sound scenes}
\acrodef{mse}[MSE]{Mean Squared Error}
\acrodef{lt}[LT]{Long-Term}
\acrodef{st}[ST]{Short-Time}
\acrodef{ism}[ISM]{Image Source Method}
\acrodef{nmse}[NMSE]{Normalized Mean Squared Error}
\acrodef{wsj0}[WSJ0]{Wall Street Journal}
\acrodef{jnd}[JND]{Just Noticeable Difference}
\acrodef{rir}[RIR]{Room Impulse Response}
\acrodef{ls}[LS]{Least Squares}
\acrodef{magls}[MagLS]{Magnitude Least Squares}
\acrodef{wdo}[WDO]{W-Disjoint Orthogonality}
\acrodef{mushra}[MUSHRA]{MUltiple Stimuli with Hidden Reference and Anchor}

\begin{document}

\begin{frontmatter}

\title{Insight into Parametric First-Order Ambisonics Upscaling in Time and Frequency Domains}

\author[bgu]{Oded Levi\corref{cor1}}
\ead{odedgeo@post.bgu.ac.il}
\author[bgu]{Boaz Rafaely}
\ead{br@bgu.ac.il}
\cortext[cor1]{Corresponding author.}

\affiliation[bgu]{organization={School of Electrical and Computer Engineering, Ben-Gurion University of the Negev},
            addressline={P.O. Box 653},
            city={Beer-Sheva},
            postcode={8410501},
            country={Israel}}

\begin{abstract}
First-Order Ambisonics (FOA) offers a practical representation for spatial audio, but its low-order reproduction introduces distinct perceptual limitations during binaural rendering. While recent methods synthesize Higher-Order Ambisonics (HOA) from FOA to overcome these constraints, the importance and potential of early reflections in the upscaling process have not yet been comprehensively explored. To bridge this knowledge gap and understand the potential of incorporating early reflections in FOA upscaling, this work develops an experimental framework that selectively upscales these reflections. By utilizing an energy-based criterion, the proposed framework identifies and selectively renders the strongest early reflections with high-order resolution. The selection mechanism is evaluated across three domains: long-term temporal, short-time temporal, and the joint time-frequency (TF) domain. Validation through objective metrics, including the Scale-Invariant Signal-to-Distortion Ratio (SI-SDR), the clarity index ($C_{50}$), and Interaural Cross-Correlation (IACC), alongside listening tests, demonstrates that the TF-domain provides superior performance. By exploiting the sparsity of speech, the TF approach maintains robust performance in complex multi-talker scenarios, providing insights into the benefits of early reflections for the design of future parametric and data-driven FOA upscaling methods.
\end{abstract}

\begin{keyword}
Ambisonics upscaling \sep binaural rendering \sep early reflections \sep First-Order Ambisonics (FOA) \sep Head-Related Transfer Function (HRTF) \sep speech sparsity \sep time-frequency processing
\end{keyword}

\end{frontmatter}

\section{Introduction}
\label{sec:introduction}

Spatial audio reproduction has gained significant attention in recent years, driven by the rapidly growing demand for highly immersive environments in virtual reality, teleconferencing, and interactive entertainment, where binaural rendering serves as a primary tool for headphone-based delivery. To encode and reproduce complex three-dimensional sound fields, Ambisonics, a framework based on the spherical harmonic decomposition of the sound field, has become a widely adopted standard \cite{rafaely2015fundamentals, zotter2019ambisonics}. Within this framework, \ac{foa} offers a practical baseline by enabling spatial capture using only a limited number of microphones \cite{zotter2019ambisonics}. However, such low-order reproduction introduces distinct acoustic and perceptual limitations; the localization of directional sounds may degrade \cite{bertet2013investigation}, audible timbral coloration may occur at high frequencies due to spectral distortion \cite{solvang2008spectral, bernschutz2014binaural}, and the captured reverberant sound field may suffer from degraded spaciousness and compromised perception \cite{avni2013spatial}.

To overcome these low-order impairments, \ac{hoa} can be employed, as increasing the spherical harmonic expansion order sharpens the directional resolution, minimizes spectral coloration, and restores natural acoustic spaciousness \cite{bertet2013investigation}. However, capturing a \ac{hoa} sound field introduces a hardware challenge; it strictly requires specialized spherical arrays composed of a high number of microphones \cite{dziwis2019development}. This extensive hardware dependency renders direct \ac{hoa} recording impractical for consumer applications and highly constrained mobile or wearable device form-factors, leaving a critical barrier to high-resolution spatial audio accessibility.

To bypass these hardware constraints while maintaining spatial resolution, parametric spatial audio methods (such as \ac{dirac} and \ac{compass}) have been proposed to synthesize \ac{hoa} from readily available \ac{foa} signals \cite{pulkki2007spatial, politis2018compass}. These synthesis frameworks operate under parametric constraints, typically assuming a limited number of dominant directional components per time-frequency bin. In complex acoustic scenes characterized by continuous multi-talker overlap, dense reverberation, or transient diffuse fields, these spatial assumptions may be violated \cite{politis2015sector}. Consequently, the underlying models focus primarily on enhancing the direct sound from sound sources rather than early reflections or late reverberation. While this approach is useful in enhancing information delivered by the direct sound \cite{miller2024role}, it fails to exploit the information in early reflections which also contributes to spatial perception \cite{zaunschirm2018binaural, laitinen2011reproducing, laitinen2012utilizing}. Furthermore, recent deep-learning-based Ambisonics upscaling approaches extending FOA to HOA \cite{routray2019deep,nawfal2025ambisonics,milstein2025diffau}, although not suffering from explicit parametric constraints, were not yet studied comprehensively and also did not provide any insight into the importance of early reflections in the upscaling process.

To address the limitations of direct-sound-focused methods, alternative approaches to spatial audio rendering have focused on the explicit reconstruction of early reflections. Reconstructing early reflections provides critical acoustic information that contributes to spatial perception, localization accuracy, and natural externalization in reverberant environments \cite{begault2001direct, catic2015role}. To capture early-reflection data, various techniques have been developed, such as blind localization methods like PHALCOR (Phase Aligned Correlation) for estimating direction-of-arrival and delays from array recordings \cite{rafaely2021blind}, and dedicated spatial processing schemes designed for spherical microphone arrays \cite{tervo2015direction}. 
However, while estimating these reflections is beneficial, current research has not yet explored the combination of explicit early reflection reconstruction with the emerging \ac{foa} enhancement techniques \cite{routray2019deep,nawfal2025ambisonics,milstein2025diffau}. Consequently, a knowledge gap remains in understanding the importance of early reflections in FOA upscaling approaches. Furthermore, it also remains unclear which processing domain, i.e. time, frequency, or a joint time-frequency, provides the preferred domain for identifying and reconstructing early-reflections information for reverberant speech signals. 

This work aims to bridge the gap identified above, by developing an analytical framework for early-reflection enhancement of FOA signals. By utilizing an energy-based guide, we identify and selectively render with high-order only the strongest early reflections within a given time, frequency, or time-frequency bin. Numerical analysis and listening test show the potential benefit of early reflections in FOA upscaling. 
The primary contributions of this paper are as follows:
\begin{itemize}
    \item \textbf{Insight into \ac{foa} upscaling:} We provide an insight into the importance of early-reflections in FOA upscaling. This insight can be useful in the design of parametric or neural FOA upscaling methods.
    \item \textbf{Framework for the selection of early reflections:} We study and provide a framework for selecting important early reflection based on sparsity and energy criterion, across three selection domains - global, temporal, and \ac{tf}, demonstrating that \ac{tf}-domain provides superior performance.
    \item \textbf{Comprehensive numerical and perceptual Validation:} We validate the proposed framework through objective metrics such as \ac{sdr}, $C_{50}$, and $\text{IACC}$), and listening tests across diverse acoustic conditions.
\end{itemize}

\section{Background}
\label{sec:background}
This section outlines the mathematical foundations for this paper. First, the formulation of the Ambisonics signal model is introduced. Subsequently, the binaural signal model is presented, integrating \ac{hrtf} coefficients in the \ac{sh} domain with Ambisonics signals. Finally, the signal models representing sound in the time and \ac{tf} domains are derived as the basis for the methodology in Sec. \ref{sec:processing_methods}.

\subsection{Ambisonics Signal Model}
Ambisonics provides a spatial audio representation that encodes a three-dimensional sound field through decomposition into \ac{sh} basis functions, in a way that is independent of the recording device. The Ambisonics signal, $a_{nm}(k)$, represents the coefficients when decomposing the plane-wave amplitude density function around the origin, $a(k, \Omega)$, into the \ac{sh} basis \cite{zotter2019ambisonics}:
\begin{equation}
    \label{eq:ambisonics_def}
    a(k, \Omega) = \sum_{n=0}^\infty \sum_{m=-n}^n a_{nm}(k)\,Y_n^m(\Omega), 
\end{equation}
where $k = 2\pi f / c$ denotes the wave number, $f$ is the frequency, $c$ is the speed of sound, 
$\Omega = (\theta, \phi)$ is the plane-wave direction of arrival, with $\theta,\phi,$ the inclination and azimuth directions, respectively, and $Y_n^m(\Omega)$ the spherical harmonic function of order $n$ and degree $m$.

Assuming the acoustic scene comprises $I$ plane waves, each representing either the direct sound from a source or a room reflection, with the i'th plane wave carrying a signal of $s_i(k)$ the Ambisonics is given by:
\begin{equation}
    \label{eq:ambisonics_freq}
    a_{nm}(k) = \sum_{i=1}^{I} s_i(k) \left[ Y_n^m(\Omega_i) \right]^*,
\end{equation}
with $\left[ \cdot \right]^{*}$ denoting the complex conjugate operator, and $\Omega_i$ the plane-wave arrival direction.

The spatial resolution captured by the Ambisonics signal is determined by the \ac{sh} order, $N$, which replaces the infinite upper limit in Eq. (\ref{eq:ambisonics_def}). When the order is truncated to $N=1$, the representation consists of four channels and is commonly referred to as \ac{foa}. Formats with truncation orders of $N > 1$ provide higher spatial resolution and are referred to as Higher-Order Ambisonics (HOA). Both representations are utilized in this work.

\subsection{Binaural Signal Model}
A binaural signal represents the acoustic pressure at the listener's ears. Computing binaural signals from Ambisonics signals requires the \acp{hrtf}, which encapsulate the direction-dependent acoustic transfer function from a sound source to the ears \cite{zotter2019ambisonics}. With the plane wave density function representing source signals, and the \ac{hrtf} the acoustic transfer function, the binaural signals, $b^{L/R}(k)$, can be computed as:
\begin{equation}
    \label{e7}
    b^{L/R}(k) = \int_{S^2} a(k, \Omega) h^{L/R}(k, \Omega) \mathrm{d}\Omega,
\end{equation}
where $h^{L/R}(k, \Omega)$ denotes the transfer function from far-field source of direction $\Omega$ to the left and right ear, respectively, and $S^2$ covers the unit sphere. 

Alternatively, Rafaely and Avni \cite{rafaely2010interaural} showed that the binaural signal can be directly calculated in the \ac{sh} domain using Eq. (\ref{eq:ambisonics_def}) and a similar equation for the \ac{hrtf} together with Parseval's theorem on the sphere, as:
\begin{equation}
    \label{eq:binaural_parseval_sh_domain}
    b^{L/R}(k) = \sum_{n=0}^{\infty} \sum_{m=-n}^{n} \left[ \tilde{a}_{nm}(k) \right]^* h_{nm}^{L/R}(k)
\end{equation}
where 

\begin{equation}
    \label{eq:a_tilde_relation}
    \tilde{a}_{nm}(k) = (-1)^m \left[a_{n,-m}(k)\right]^*
\end{equation}
is the spherical Fourier transform of the complex conjugate of the amplitude density function, and $h_{nm}^{L/R}(k)$ represents the \ac{sh} coefficients of the \ac{hrtf} for the respective ear.

By substituting Eq. \eqref{eq:ambisonics_freq} and Eq. \eqref{eq:a_tilde_relation} into Eq. \eqref{eq:binaural_parseval_sh_domain},  
and limiting the infinite series to order $N$, and using complex-conjugate symmetry in the \ac{sh} domain \cite{rafaely2015fundamentals}, the synthesized binaural signal for a plane-wave sound field can be formulated as:
\begin{equation}
    \label{eq:binaural_expanded}
    b^{L/R}(k) = \sum_{n=0}^{N} \sum_{m=-n}^{n} \sum_{i=1}^{I} s_i(k) Y_n^m(\Omega_i) h_{nm}^{L/R}(k) 
\end{equation}

In Eq.~\eqref{eq:binaural_expanded}, the \ac{hrtf} coefficients $h_{nm}^{L/R}(k)$ are the \ac{ls} solution of the \ac{sht} of the \ac{hrtf}, truncated to order $N$. For low orders, and in particular for \ac{foa}, this truncation leads to a loss of high-frequency energy and to audible timbral coloration \cite{bernschutz2014binaural}. The \ac{magls} decoder \cite{schorkhuber2018binaural} reduces this impairment by fitting, above a cutoff frequency, only the magnitude of the \ac{hrtf}. In this work, \ac{foa} signals are rendered with the order-1 \ac{magls} decoder, while high-order components are rendered with the \ac{ls} decoder.

\subsection{Signal Model in the Time and Time-Frequency Domains}
\label{sec:time_tf_models}
The proposed method operates directly on the binaural signals in the time and joint \ac{tf} domains, and so formulation in these domains is presented next.
The continuous-time binaural signals, $b^{L/R}(t)$, are obtained by applying the \ac{ift} to the frequency-domain binaural signals $b^{L/R}(k)$:
\begin{equation}
    \label{eq:binaural_time}
    b^{L/R}(t) = \mathcal{F}^{-1} \{ b^{L/R}(k) \}
\end{equation}
where $\mathcal{F}^{-1}\{\cdot\}$ denotes the \ac{ift} operator.

These signals are further transformed into the joint \ac{tf} domain using the \ac{stft}. The \ac{tf} representation of the binaural signals, denoted as $B^{L/R}(\tau, \nu)$, is formulated as:
\begin{equation}
    \label{eq:binaural_tf}
    B^{L/R}(\tau, \nu) = \mathcal{S} \{ b^{L/R}(t) \} 
\end{equation}
where $\tau$ and $\nu$ denote the time-frame and discrete frequency indices, respectively, and $\mathcal{S}\{\cdot\}$ represents the \ac{stft} operator.

\section{Problem Formulation}
\label{sec:problem_formulation}

While Ambisonics has been accepted as a principal spatial audio format \cite{zotter2019ambisonics}, \ac{foa} in particular has been widely adopted due to its small channel count and the availability of recording hardware \cite{Zhang2017review}. However, its low spatial resolution inherently limits the perceived spatial quality \cite{avni2013spatial}. Conversely, extending the order of \ac{foa} is expected to have a direct effect on spatial perception quality. 
To address this limitation, various methods have been proposed to enhance or upscale \ac{foa} signals. These include parametric spatial audio techniques such as \ac{dirac} \cite{pulkki2007spatial} and \ac{compass} \cite{politis2018compass}, as well as more recent data-driven Ambisonics upscaling approaches utilizing deep learning \cite{routray2019deep,nawfal2025ambisonics,milstein2025diffau}. 

While an upscaling of \ac{foa} is highly desirable, current literature presents only initial studies, which may benefit from some insight into the upscaling problem. 
In particular, it may be beneficial to focus on perceptually dominant sound field components. Recent studies have shown that extending the spatial order of the direct sound \cite{miller2024role} and prominent early reflections \cite{lubeck2020perception, engel2022reverberation} improves aspects of spatial spatial perception.

Despite this understanding, a comprehensive investigation into exactly which reflections or components should be upscaled was not presented, and a selection methodology for these components has not been proposed. 

Therefore, the problem addressed in this work is defined as follows: Given a \ac{foa} representation, what is the preferred framework for selecting a limited subset of room reflections to be upscaled to \ac{hoa} with the aim of improving spatial perception? 

To formulate this problem mathematically, consider Eq. (\ref{eq:binaural_expanded}), in which $I$ represents the direct sound and all room reflections, and the Ambisonics signal is separated into \ac{foa} and the remaining higher orders:

\begin{align}
    \label{eq:binaural_decomposed}
    b^{L/R}(k) &= \sum_{n=0}^{1} \sum_{m=-n}^{n} \sum_{i=1}^{I} s_i(k) Y_n^m(\Omega_i) h_{nm}^{L/R}(k) \nonumber \\
    &+ \sum_{n=2}^{N} \sum_{m=-n}^{n} \sum_{i\in\Psi} s_i(k) Y_n^m(\Omega_i) h_{nm}^{L/R}(k),
\end{align}
where $\Psi$ is a selected subset of room reflections. Rewritten in the time domain, and evaluated over the \ac{sh}, the decomposition into the two binaural signal components can be written as:

\begin{align}
    \label{eq:binaural_decomposed_time}
    b^{L/R}(t) &= \sum_{i=1}^{I} b_{FOA,i}^{L/R}(t) + \sum_{i\in\Psi} b_{\Psi,i}^{L/R}(t)\nonumber \\
               &= b_{FOA}^{L/R}(t) + b_{\Psi}^{L/R}(t).
\end{align}

With \ac{ls} decoding, $b_{\Psi,i}^{L/R}(t)$ is the contribution of orders $2,\dots,N$ of the $i$-th reflection, as in Eq.~\eqref{eq:binaural_decomposed}. When the \ac{foa} component is rendered with \ac{magls}, the upscaled component is defined more generally as the difference between the order-$N$ \ac{ls} rendering of the $i$-th reflection and its \ac{foa} rendering, $b_{\Psi,i}^{L/R}(t) = b_{HOA,i}^{L/R}(t) - b_{FOA,i}^{L/R}(t)$, so that each selected reflection is rendered exactly as in the \ac{hoa} reference, while all other reflections remain in \ac{foa}.

The next section presents the proposed method, where a selection criterion for $\Psi$ and the reconstruction of $b_{\Psi}^{L/R}(t)$ are presented and investigated when the selection is performed over the entire signal, over individual time frames and in \ac{tf}.

\section{Processing Methods}
\label{sec:processing_methods}

This section presents the proposed methodology for selecting the subset of room reflections, $\Psi$, to be upscaled to \ac{hoa}. As established in Sec. \ref{sec:problem_formulation}, rather than applying an upscaling to the entire sound field, the proposed approach focuses on identifying dominant components. To achieve this, the selection criterion evaluates the energy of the \ac{foa} binaural signal components, motivated by the approach utilized in \ac{dirac} \cite{pulkki2007spatial}. 

By analyzing for each reflection the combined left and right ear of the \ac{foa} components after decoding to binaural signals (with the order-1 \ac{ls} decoder), the contribution of each reflection can be estimated. Once the subset $\Psi$ containing $r$ reflections is determined through this criterion, their corresponding higher-order components are synthesized to form $b_{\Psi}^{L/R}(t)$. This signal is then added to the base \ac{foa} signal $b_{FOA}^{L/R}(t)$, following Eq. \eqref{eq:binaural_decomposed_time}. The subset size $r$ serves as a tunable hyperparameter, controlling the trade-off between accuracy and complexity. We investigate this selection criteria across three distinct processing methods: Long-term, Short-time, and Time-Frequency (\ac{tf}).

\subsection{Long-term Processing}
In the \ac{lt} approach, the energetic contribution of each reflection is evaluated over the entire duration of the signal. The long-term energy for the $i$-th reflection is defined, following \eqref{eq:binaural_decomposed_time}, as the sum of the squared amplitudes of its left and right binaural signals across all discrete time samples $t$:

\begin{equation}
    \label{eq:energy_longterm}
    E_{\text{LT}}(i) = \sum_{t} \Big[ \big( b_{\text{FOA}, i}^L(t) \big)^2 + \big( b_{\text{FOA}, i}^R(t) \big)^2 \Big].
\end{equation}

Once $E_{\text{LT}}(i)$ is computed for all $I$ reflections, the subset $\Psi_{\text{LT}}$ is formed by selecting the $r$ reflections that maximize the total energy:
\begin{equation}
    \label{eq:subset_longterm}
    \Psi_{\text{LT}} = \underset{\Psi \subset \{1, \dots, I\}, \, |\Psi| = r}{\mathrm{arg\,max}} \sum_{i \in \Psi} E_{\text{LT}}(i) .
\end{equation}

Because this selection is static, the same subset $\Psi_{\text{LT}}$ is utilized throughout the entire signal duration. The final approximated signal is reconstructed directly by substituting $\Psi_{\text{LT}}$ into Eq. \eqref{eq:binaural_decomposed_time}.

\subsection{Short-time Processing}
While the Long-term method provides a static approximation, acoustic scenes and speech signals exhibit dynamic temporal variations. To account for these fluctuations, the \ac{st} processing evaluates the energy within localized, overlapping time frames. 

The signals are segmented using a window function $w(t)$ with a predefined frame length and step size that satisfy the overlap-add constraints \cite{crochiere1980weighted}. Let $l$ denote the frame index and $\Delta$ represent the step size. The energy for each reflection $i$ at frame $l$ is computed as:
\begin{align}
    E_{\text{ST}}(i, l) &= \sum_{t} \Big[ \left( b_{FOA, i}^L(t) \cdot w(t - l\Delta)\right) ^2 \notag \\
    &\quad + \left(b_{FOA, i}^R(t) \cdot w(t - l\Delta)\right)^2 \Big] .
    \label{eq:energy_shorttime}
\end{align}

For each frame index $l$, the subset $\Psi_{\text{ST}}(l)$ is determined by solving:
\begin{equation}
    \label{eq:subset_shorttime}
    \Psi_{\text{ST}}(l) = \underset{\Psi \subset \{1, \dots, I\}, \, |\Psi| = r}{\mathrm{arg\,max}} \sum_{i \in \Psi} E_{\text{ST}}(i, l) .
\end{equation}

The higher-order signals of the selected reflections in $\Psi_{\text{ST}}(l)$ are windowed and summed for each frame to form the localized residual component, and the time-domain signal $b_{\Psi}^{L/R}(t)$ is reconstructed using the overlap-add procedure \cite{crochiere1980weighted}.

\subsection{Time-Frequency Processing}
Speech signals exhibit a high degree of sparsity in the \ac{tf} domain. Due to the W-Disjoint Orthogonality property \cite{yilmaz2004blindseparation}, multiple competing speech signals mostly do not overlap in \ac{tf}. Therefore, the benefit of a \ac{tf} selection is expected mainly in multi-talker scenes. Furthermore, human auditory spatial perception operates on a frequency-band basis \cite{faller2003binauralcue}. Consequently, adapting the selection criterion to the \ac{tf} domain allows the framework to exploit both the sparsity of speech and the frequency-dependent nature of human spatial perception.

In this approach, the binaural signals are transformed into the \ac{tf} domain using the \ac{stft}, resulting in the complex representations $B_{FOA, i}^{L/R}(\tau, \nu)$, as in Eq. (\ref{eq:binaural_tf}).
The energy in this case is computed over a localized region, defining a rectangular window $\mathcal{W}(\tau, \nu)$ of dimensions $\Delta_\tau \times \Delta_\nu$ centered around $(\tau, \nu)$. The energy of the $i$-th reflection is therefore computed as:
\begin{align}
    E_{\text{TF}}(i, \tau, \nu) &= \sum_{(\tau', \nu') \in \mathcal{W}(\tau, \nu)} \Big( \left|B_{FOA, i}^L(\tau', \nu')\right|^2 \notag \\
    &\quad + \left|B_{FOA, i}^R(\tau', \nu')\right|^2 \Big) .
    \label{eq:energy_tf}
\end{align}

Based on this energy, for every central bin $(\tau, \nu)$, the optimal subset of reflections is determined by maximizing the energy as before:
\begin{equation}
    \label{eq:subset_tf}
    \Psi_{\text{TF}}(\tau, \nu) = \underset{\Psi \subset \{1, \dots, I\}, \, |\Psi| = r}{\mathrm{arg\,max}} \sum_{i \in \Psi} E_{\text{TF}}(i, \tau, \nu) .
\end{equation}

The two components of the binaural signal are then added, as in Eq. (\ref{eq:binaural_decomposed_time}), only in the \ac{tf} domain, after which the signal is transformed back to the time domain by applying the inverse \ac{stft} \cite{allen1977unified}.

\section{Performance Measures} 
\label{sec:measures}

To evaluate the proposed binaural rendering framework, a set of performance measures quantifying binaural error, signal distortion, and reverberation preservation is employed. The approximated binaural signal, $\hat{b}^{L/R}(t)$, obtained via Eq. \eqref{eq:binaural_decomposed_time} with the reflection subset $\Psi$ selected according to the \ac{lt}, \ac{st}, or \ac{tf} criteria detailed in Sec. \ref{sec:processing_methods}, is compared against a reference binaural signal, $b_{\text{ref}}^{L/R}(t)$, obtained by rendering the full \ac{hoa} representation, where $\Psi = \{1,\dots,I\}$, i.e. includes all reflections. 

To support a more detailed acoustic analysis, both the reference and the approximated signals are further decomposed into a direct-sound component and a reverberant component,
\begin{equation}
    \label{eq:binaural_partition}
    b^{L/R}(t) = b_{\text{direct}}^{L/R}(t) + b_{\text{reverb}}^{L/R}(t).
\end{equation}

The performance measures are detailed below:

\begin{itemize}
    
    \item \textbf{Scale-Invariant Signal-to-Distortion Ratio (SI-SDR):} Measures signal distortion while ignoring global amplitude differences \cite{leroux2019sdr}. To avoid introducing interaural bias and distorting spatial cues (e.g., interaural level differences), the optimal scaling factor $\alpha$ is computed jointly over both channels $c \in \{L, R\}$ rather than independently, so that any differences between channels are not normalized by the measure. The global target scaling is defined via orthogonal projection as $\alpha = \frac{\sum_c \sum_t \hat{b}^c(t) b_{\text{ref}}^c(t)}{\sum_c \sum_t (b_{\text{ref}}^c(t))^2}$. The overall SI-SDR is then evaluated as:
    \begin{equation}
        \label{eq:sdr}
        \text{SI-SDR} = 10 \log_{10} \left( \frac{ \sum\limits_{c \in \{L, R\}} \sum\limits_t \left( \alpha b_{\text{ref}}^c(t) \right)^2 }{ \sum\limits_{c \in \{L, R\}} \sum\limits_t \left( \hat{b}^c(t) - \alpha b_{\text{ref}}^c(t) \right)^2 } \right) .
    \end{equation}
    This metric is computed for the complete binaural signal, the direct sound component, and the direct sound together with the early reflections.

    \item \textbf{Clarity Index Difference ($\Delta C_{50}$):} 
    Evaluates the preservation of speech clarity by measuring the ratio of early sound energy (arriving within the first 50~ms after the direct sound) to the late reverberant energy. The $C_{50}$ of a binaural signal $b^{L/R}(t)$ is defined as:
    \begin{equation}
        \label{eq:c50}
        C_{50} = 10 \log_{10} \left( \frac{\sum\limits_{t \le 50\text{ ms}} \left[ \left( b^L(t) \right)^2 + \left( b^R(t) \right)^2 \right]}{\sum\limits_{t > 50\text{ ms}} \left[ \left( b^L(t) \right)^2 + \left( b^R(t) \right)^2 \right]} \right) .
    \end{equation}
     Evaluating Eq. \eqref{eq:c50} for the reference signal and the approximated signal yields $C_{50,\text{ref}}$ and $C_{50,\text{est}}$, respectively. The approximation error is evaluated as
    \[\Delta C_{50} = \left| C_{50,\text{ref}} - C_{50,\text{est}} \right|.\]

    \item \textbf{Interaural Cross-Correlation Difference ($\Delta$IACC):} 
    The \ac{iacc} of a binaural signal $b^{L/R}(t)$ is defined as the maximum cross-correlation coefficient between the left and right ears within a plausible human interaural time difference limit of $|\tau| \le 1\text{ ms}$:
    \begin{equation}
        \label{eq:iacc}
        \text{IACC} = \max_{|\tau| \le 1\text{ ms}} \frac{\sum\limits_t b^L(t) b^R(t+\tau)}{\sqrt{\sum\limits_t \left(b^L(t)\right)^2 \sum\limits_t \left(b^R(t)\right)^2}} .
    \end{equation}
    Evaluating Eq. \eqref{eq:iacc} for the reference signal $b_{\text{ref}}^{L/R}(t)$ and for the approximated signal $\hat{b}^{L/R}(t)$ yields $\text{IACC}_{\text{ref}}$ and $\text{IACC}_{\text{est}}$, respectively, and the deviation is quantified as the absolute error
    \[\Delta\text{IACC} = \left| \text{IACC}_{\text{ref}} - \text{IACC}_{\text{est}}\right|.\]
    When averaged over scenes, $\Delta$IACC is reported as the root-mean-square value.
\end{itemize}

\section{Simulation Study}
\label{sec:simulation}

To evaluate the performance of the proposed framework across the three investigated processing domains (\ac{lt}, \ac{st}, and \ac{tf}), a comprehensive simulation study was conducted. The following subsections detail the acoustic simulation environments, the applied methodology, and present a discussion of the experimental results based on the evaluation metrics defined in Sec. \ref{sec:measures}.

\subsection{Experimental Setup}
\label{subsec:setup}

The acoustic environments were simulated using the \ac{ism} \cite{allen1979image} to model six distinct shoebox-shaped rooms of increasing volumes. The room dimensions and their corresponding reverberation times ($T_{60}$) were chosen to represent diverse acoustic conditions. Specifically, the $T_{60}$ values were increased from $0.3\,$s for the small room of size $6 \times 4 \times 3$~m, to $0.6$~s for the fourth room size-wise in uniform increments of $0.1$~s per room, while the two largest rooms (with the largest of size $20 \times 15 \times 6$~m), have a $T_{60}$ of $0.8$~s and $0.65$~s. A comprehensive summary of the experimental parameters and ranges is provided in Table~\ref{tab:room_setup}.

\begin{table}[htbp]
    \centering
    \caption{Details of the simulated acoustic environments and ranges of parameters.}
    \label{tab:room_setup}
    \renewcommand{\arraystretch}{1.2}
    \resizebox{\columnwidth}{!}{%
    \begin{tabular}{@{}l >{\raggedright\arraybackslash}p{4.8cm}@{}} 
    \toprule
    \textbf{Parameter} & \textbf{Value / Description} \\
    \midrule
    Number of Rooms & $6$ with volumes $72$--$1800$~m$^3$ \\
    Rooms Dimensions ($L \times W \times H$) & $[6.0, 20.0] \times [4.0, 15.0] \times [3.0, 6.0]$~m \\
    $T_{60}$ Range & $[0.3, 0.8]\,$s \\
    Source Positions & $3$ positions per room ($\ge$ $0.5$~m from boundaries) \\
    Array Positions & $10$ varying distances per source position ($30$ total per room) \\
    Direct-to-Reverberant Ratio & $[-10, 15]$~dB \\
    Number of Sources & $\sim \mathcal{U}\{1, 4\}$ concurrent speakers \\
    Speech Source & $3.4\,$s clean speech from the \ac{wsj0} corpus \\
    \bottomrule
    \end{tabular}%
    }
\end{table}

In each room, the listener was placed at $(0.31L, 0.31W, 1.7)$~m, facing the $x$ axis. Within each of the six rooms, $30$ distinct acoustic scenes were evaluated by defining three source directions, at azimuths of $30^\circ$, $60^\circ$ and $75^\circ$ in the horizontal plane at head height. For each direction, the source distance was varied across $10$ discrete points with geometric spacing. These distances were designed to cover a wide range of acoustic clarity, from a \ac{drr} of $-10$~dB at the far end to $+15$~dB at the near end, where the \ac{drr} is the ratio between the energy of the direct sound and that of all other image sources. Where a room does not reach these values, the distance was limited to $0.5$~m from the head and $0.5$~m from the walls. Therefore, the high values are reached only in the larger rooms, whereas in the smaller room the maximum \ac{drr} is $5.5$~dB.

For each acoustic scene, the number of active sound sources was drawn from a discrete uniform distribution $\mathcal{U}\{1, 4\}$. For single talker scenes, the talker was placed at the source directions of $30^\circ$, $60^\circ$ or $75^\circ$, while multiple talkers were placed on both sides of the frontal direction, at azimuths of $\pm 75^\circ$, $\{0^\circ, \pm 75^\circ\}$ and $\{\pm 25^\circ, \pm 75^\circ\}$ for two, three and four talkers, respectively, all at the same distance and level. For multi-talker scenes, the \ac{drr} was computed for each talker and averaged. The source signals consisted of $3.4$-second clean, anechoic speech excerpts randomly selected from the \ac{wsj0} corpus \cite{paul1992design, garofolo1993csr}, with at least one male and one female talker in every multi-talker scene.

To isolate the performance of the proposed selection algorithm from spatial aliasing and encoding artifacts inherent to physical microphone arrays, the sound field was modeled directly in the \ac{sh} domain. Specifically, ideal Ambisonics signals were generated directly using the \ac{ism} up to a truncation order of $N = 10$. The general formulation for this \ac{sh} decomposition was introduced earlier in Eq.~\eqref{eq:binaural_decomposed}. Binaural signals were subsequently synthesized from these ideal \ac{sh} coefficients using a high-resolution \ac{hrtf} dataset measured from a Neumann KU100 dummy head, sampled on a Lebedev grid \cite{bernschutz2013spherical, lebedev1976quadratures}.

\subsection{Methodology}
\label{subsec:methodology}

To evaluate the processing methods described in Sec.~\ref{sec:processing_methods}, the acoustic paths of each active source were partitioned into two components: (1) the direct sound and early reflections arriving within $100\,$ms, and (2) the late reverberation. The combined direct sound and early reflections across all active sources comprise the set of $I$ candidate reflections, as defined in Sec.~\ref{sec:problem_formulation}.

The selection criterion operates on the \ac{foa} components of these candidates. Late reverberation is excluded from the selection process and is rendered in \ac{foa}. From the $I$ candidate reflections, a subset $\Psi$ of size $r \in \{0, 1, 3, 10, 30\}$ is selected for \ac{hoa} rendering, following Eq.~\eqref{eq:binaural_decomposed} and Eq.~\eqref{eq:binaural_decomposed_time}. The selection mechanism is evaluated across three domains:

\begin{enumerate}
    \item \textbf{Long term (\ac{lt}):} Selection is based on the global energy computed over the entire signal duration.
    \item \textbf{Short-time (\ac{st}):} Signals are segmented using a periodic Hann window of $32\,$ms. The temporal step size $\Delta$, defined in Eq.~\eqref{eq:energy_shorttime}, is set to $16\,$ms (50\% overlap). The subset of $r$ reflections is selected per frame based on this localized energy.
    \item \textbf{Time-Frequency (\ac{tf}):} Using the same $32\,$ms windowing, the energy is computed over a localized rectangular region $\mathcal{W}(\tau, \nu)$, as defined in Eq.~\eqref{eq:energy_tf}, with dimensions $\Delta_\tau = 3$ temporal frames and $\Delta_\nu = 3$ frequency bins. Based on this smoothed energy, the subset of $r$ reflections is determined for each time-frequency bin.
\end{enumerate}

To evaluate the proposed framework, the processed signals were compared with a reference signal. This reference was synthesized by rendering all $I$ candidate reflections in full \ac{hoa}, together with the late reverberation. Using this reference signal, the performance was quantified through the evaluation metrics defined in Sec.~\ref{sec:measures}. Specifically, the overall reconstruction quality was assessed using the \ac{sdr} metric defined in Eq.~\eqref{eq:sdr}, which was evaluated across three distinct signal partitions:
\begin{itemize}
    \item \text{\acs{sdr} - Direct :} The direct sound component exclusively.
    \item \text{\acs{sdr} - Direct+Early :} The direct sound combined with the early reflections.
    \item \text{\acs{sdr} - Global :} The complete signal, comprising the direct sound, early reflections, and late reverberation.
\end{itemize}
Furthermore, the spatial preservation and signal clarity were evaluated using the $\Delta$\ac{iacc} and $\Delta C_{50}$ metrics, respectively.

\subsection{Results and Discussion}
\label{subsec:results}

The proposed framework was evaluated across the simulated acoustic environments detailed in Sec.~\ref{subsec:setup}. The performance analysis is structured into three main parts. First, the overall reconstruction accuracy is established across all acoustic scenes. Next, the preservation of key perceptual and spatial attributes is evaluated. Finally, the robustness is assessed under varying acoustic complexities, specifically examining the impact of different reverberation conditions and multiple speakers.

\subsubsection{Overall Reconstruction Accuracy}
\begin{figure}[htbp]
    \centering
    \includegraphics{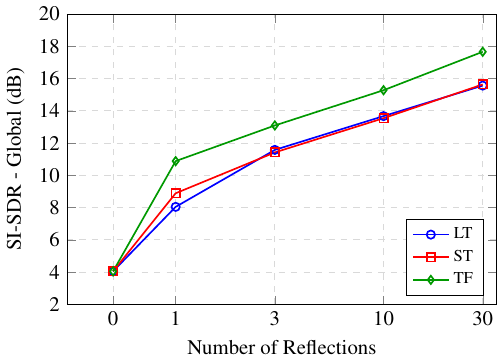}
    \caption{Global \acs{sdr} performance evaluated across all acoustic scenes.}
    \label{fig:SDR_Global}
\end{figure}

\begin{figure}[htbp]
    \centering
    \includegraphics{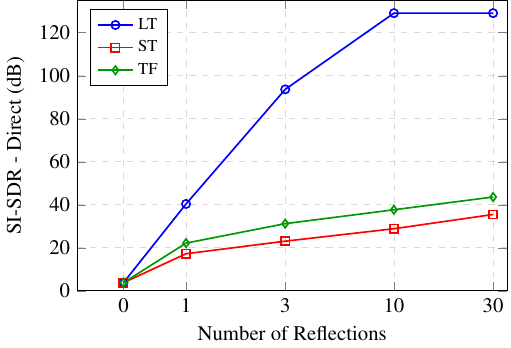}
    \caption{\acs{sdr} evaluating the reconstruction of the direct signal components.}
    \label{fig:SDR_Direct}
\end{figure}

\begin{figure}[htbp]
    \centering
    \includegraphics{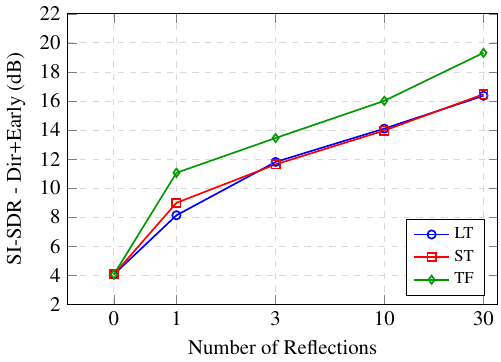}
    \caption{\acs{sdr} evaluating the reconstruction of direct and early reflection components.}
    \label{fig:SDR_Direct_Early}
\end{figure}

This section evaluates the overall reconstruction accuracy of the proposed methods and examines how the estimation quality is distributed between the direct sound, the early reflections, and the late reverberation. The performance is assessed using the \ac{sdr} metric, and the results as a function of the number of reflections $r$ are depicted in Figs.~\ref{fig:SDR_Global}--\ref{fig:SDR_Direct_Early}.

Fig.~\ref{fig:SDR_Global} shows the global SI-SDR, illustrating that, performance improves monotonically as the number of selected reflections increases, validating the core premise of the proposed approach. Furthermore, a notable divergence emerges as $r$ grows: while the \ac{lt} and \ac{st} baselines exhibit similar performance, the \ac{tf} method demonstrates a consistent advantage. At $r=30$, the \ac{tf} approach yields a global \acs{sdr} of 17.66~dB, outperforming both \ac{st} (15.64~dB) and \ac{lt} (15.56~dB). This relative improvement demonstrates that operating in the joint time-frequency domain exploits the sparsity of speech, allowing for a more efficient utilization of the selected subset compared with the temporal baselines. A similar behavior is shown in Fig.~\ref{fig:SDR_Direct_Early} for the SI-SDR of the direct and early reflections component.

However, an independent evaluation of the direct sound (Fig.~\ref{fig:SDR_Direct}) reveals a distinctly different behavior. At $r=30$, the \ac{lt} method achieves an exceptionally high \acs{sdr} - Direct of approximately 129.1~dB, which corresponds to an exact reconstruction, limited only by numerical precision. Because \ac{lt} selects reflections based on global energy averages, it consistently prioritizes the highly energetic direct sound over later reflections. Conversely, the \ac{st} and \ac{tf} methods yield lower direct-sound accuracy (35.50~dB and 43.62~dB, respectively), as their localized selection mechanisms distribute the allocated reflections across various temporal and frequency components rather than the direct sound. Nevertheless, this balanced selection strategy by the \ac{tf} method proves advantageous when evaluating the combined direct and early reflections (Fig.~\ref{fig:SDR_Direct_Early}), where \ac{tf} achieves 19.32~dB compared to \ac{lt}'s 16.36~dB.

\subsubsection{Preservation of $C_{50}$ and \ac{iacc}}
\begin{figure}[htbp]
    \centering
    \includegraphics{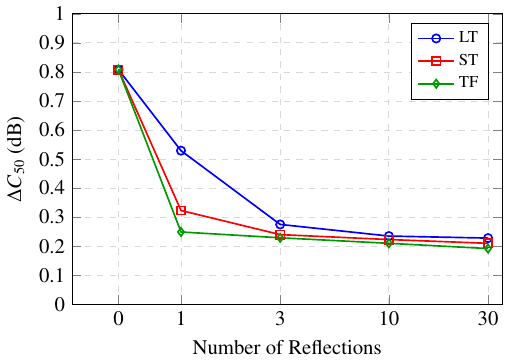}
    \caption{Absolute error in clarity index ($\Delta C_{50}$) across different reflection orders.} 
    \label{fig:Delta_C50}
\end{figure}

\begin{figure}[htbp]
    \centering
    \includegraphics{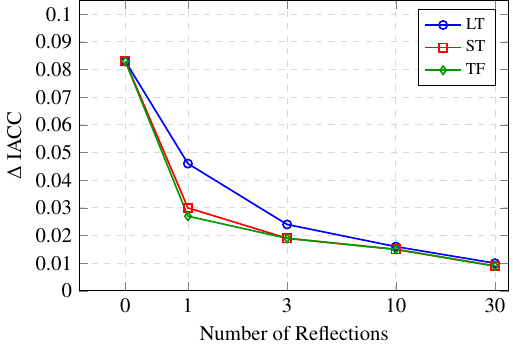}
    \caption{Error in \ac{iacc} ($\Delta$\acs{iacc}) across different reflection orders.}
    \label{fig:Delta_IACC}
\end{figure}

This section evaluates the preservation of early-to-late energy clarity ($C_{50}$) and interaural cross-correlation (\ac{iacc}). The absolute errors for these metrics as a function of the number of reflections are presented in Figs.~\ref{fig:Delta_C50} and \ref{fig:Delta_IACC}. $C_{50}$ is evaluated relative to the \ac{jnd} threshold of 1.1~dB \cite{bradley1999just}.

\begin{itemize}
    \item \textbf{Clarity Preservation ($\Delta C_{50}$):} Fig.~\ref{fig:Delta_C50} shows a steady decline in the error as the number of reflections increases. At $r=30$, the \ac{tf} method yields a $\Delta C_{50}$ of 0.19~dB, slightly lower than both \ac{st} (0.21~dB) and \ac{lt} (0.23~dB). These values fall well below the \ac{jnd}, ensuring preservation of the ratio of early to late reverberation under all conditions.
    
    \item \textbf{Interaural Cross-Correlation ($\Delta$\acs{iacc}):} Fig.~\ref{fig:Delta_IACC} illustrates a notable improvement in interaural correlation, which is evident even with the inclusion of a single reflection ($r=1$). The \ac{tf} method reduces the $\Delta$\acs{iacc} to 0.009 at $r=30$ (down from 0.083 at $r=0$). The \ac{st} and \ac{lt} baselines also show significant, similar improvements, confirming the effective preservation of binaural coherence.
\end{itemize}

\subsubsection{Robustness to Varying \ac{drr}}
To evaluate the algorithmic robustness across varying reverberation conditions, the simulated scenes were partitioned into two \ac{drr} groups. The first group covers a low-to-moderate \ac{drr} range (\ac{drr} $\in [-10, 5)$~dB), while the second group focuses on a high \ac{drr} range (\ac{drr} $\in [5,$ $15]$~dB), representing conditions dominated by the direct sound. 

\begin{figure}[htbp]
    \centering
    \includegraphics{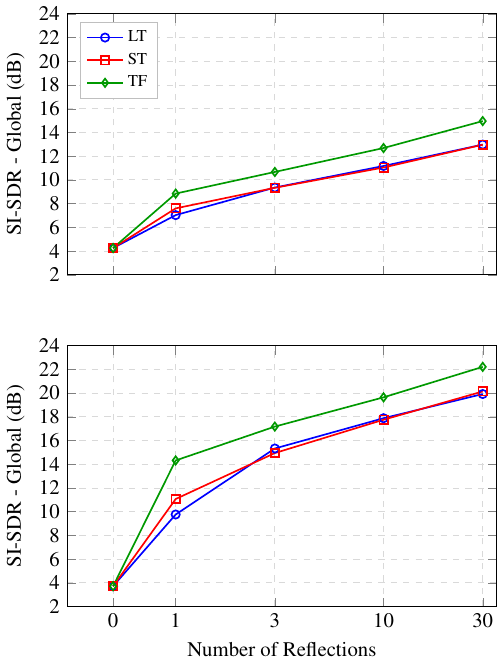}
    \caption{Global \acs{sdr} performance evaluated across two distinct DRR conditions. Top: DRR $\in$ [-10,5] dB. Bottom: DRR $\in$ [5,15] dB.}
    \label{fig:SDR_DRR_Groups}
\end{figure}

The global \ac{sdr} performance for both groups is presented in Fig.~\ref{fig:SDR_DRR_Groups}. While other metrics, such as early-to-late clarity and interaural correlation, were also evaluated under these partitioned conditions, they exhibited consistent behavioral trends. Therefore, for clarity and conciseness, the global \acs{sdr} is presented as the primary representative metric in this section.

The overall reconstruction accuracy in the more reverberant conditions (Fig.~\ref{fig:SDR_DRR_Groups}, top panel) is lower across all methods compared to the less reverberant conditions (bottom panel), due to the pronounced importance of the direct sound and early reflections. Under the reverberant conditions, the \ac{tf} method outperforms the temporal baselines, achieving an \acs{sdr} of 14.96~dB at $r=30$, compared to 12.97~dB and 12.98~dB for the \ac{st} and \ac{lt} approaches, respectively. 

In the less reverberant conditions (Fig.~\ref{fig:SDR_DRR_Groups}, bottom panel), the \ac{tf} method is able to more effectively isolate the dominant sparse speech components. At $r=30$, it reaches a high \acs{sdr} of 22.21~dB, higher than the \ac{st} (20.14~dB) and \ac{lt} (19.92~dB) methods. This consistent advantage across both subsets confirms the superiority of the joint time-frequency domain across this wide range of conditions.

\subsubsection{Target reconstruction in Multi-Talker Scenarios}
\label{subsubsec:multi}
\begin{figure}[htbp]
    \centering
    \includegraphics{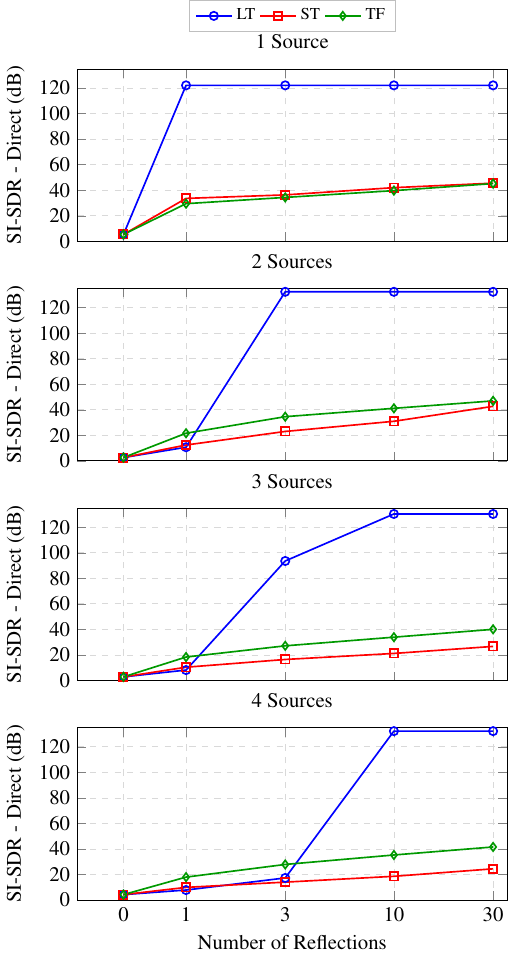}
    \caption{Performance of direct sound extraction (\acs{sdr} Direct) under multi-source scenarios (1 to 4 concurrent active speakers).}
    \label{fig:Multi_Source_SDR}
\end{figure}

\begin{figure}[htbp]
    \centering
    \includegraphics{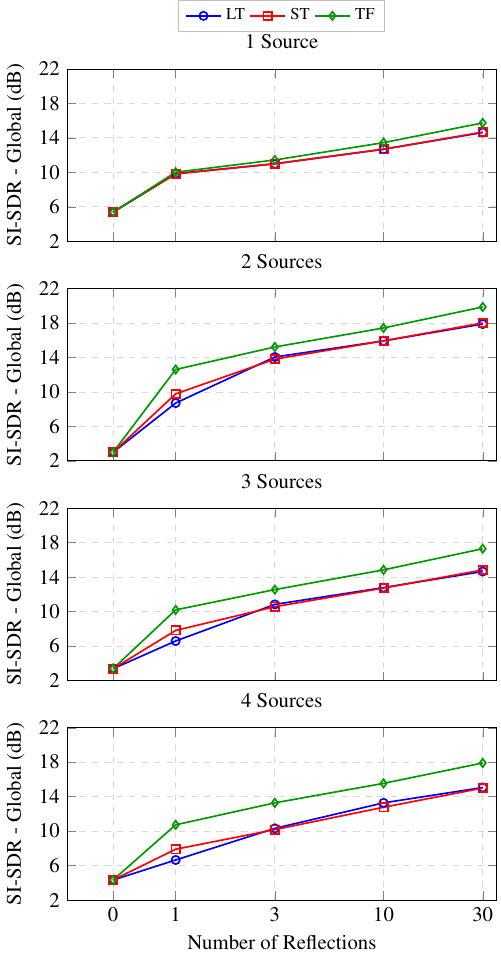}
    \caption{Global \acs{sdr} under multi-source scenarios (1 to 4 concurrent active speakers).}
    \label{fig:Multi_Source_SDR_Global}
\end{figure}

To assess the capability of the approach under study in isolating the target signal amidst multiple speakers, a multi-source analysis was conducted. This evaluation first focuses on the extraction of the direct sound (\acs{sdr} Direct) which is critical for target localization and speech intelligibility, and then on the complete signal (\acs{sdr} Global). The performance was evaluated under scenarios containing 1, 2, 3, and 4 concurrent active sources, as illustrated in Figs.~\ref{fig:Multi_Source_SDR} and \ref{fig:Multi_Source_SDR_Global}.

In the single-source scenario (Fig.~\ref{fig:Multi_Source_SDR}, top panel), both the \ac{st} and \ac{tf} methods exhibit comparable performance. However, as the number of interfering sources increases, the temporal overlap between competing signals becomes dense, leading to degradation in the \ac{st} method's performance. On the other hand, the \ac{tf} method demonstrates robustness to the increased number of speakers. By leveraging sparse time-frequency representations of reverberant speech, the \ac{tf} approach successfully isolates the disjoint target components even when multiple sources are active simultaneously. Under the most challenging condition of 4 concurrent sources (Fig.~\ref{fig:Multi_Source_SDR}, bottom panel), the \ac{tf} method achieves an \acs{sdr} of 41.60~dB at $r=30$, maintaining a notable margin over the \ac{st} method (24.48~dB). The \ac{lt} method shows an interesting behavior. When the number of reflections is larger or equal to the number of sources, it shows very high \ac{sdr} values, due to its robust temporal processing. However, as soon as the number of sources increases above the number of reflections, performance drops below that of the other methods. Overall, in multiple speaker environment, the \ac{tf} method seems preferable if the number of upscaled reflections is not large.

The global \acs{sdr} (Fig.~\ref{fig:Multi_Source_SDR_Global}) shows how these differences affect the complete signal. In the single-source scenario (top panel), the three methods exhibit almost identical performance; with a single reflection, they reach 9.82--10.02~dB. With a single talker, the direct sound is typically the dominant, so all methods seem to select it first. As soon as more than one talker is active, the \ac{tf} method can assign each \ac{tf} bin to the dominant reflection (mostly the direct sound) of the talker active in it and can therefore outperform both temporal methods in this case. The margin is largest for a single reflection, where \ac{tf} outperforms \ac{lt} by 3.9~dB for two talkers (12.60~dB compared with 8.71~dB), by 3.6~dB for three talkers and by 4.1~dB for four talkers. With more reflections, the temporal methods approach \ac{tf}: for two talkers, \ac{lt} with three reflections (14.05~dB) already exceeds \ac{tf} with a single reflection.

\begin{figure}[htbp]
    \centering
    \includegraphics{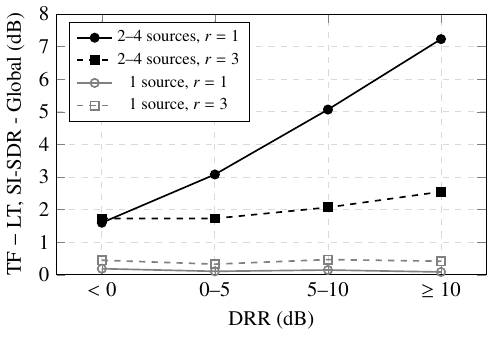}
    \caption{Improvement of \acs{tf} over \acs{lt} in global \acs{sdr} as a function of the \ac{drr}, for single-source scenes (average of 50 scenes) and multi-source scenes (average of 130 scenes).}
    \label{fig:TF_gain_DRR}
\end{figure}

The dependence of this advantage on the \ac{drr} is shown in Fig.~\ref{fig:TF_gain_DRR}, which presents the improvement of \ac{tf} over \ac{lt} in global \acs{sdr} for four \ac{drr} ranges, separately for single-talker and multi-talker scenes. For a single talker, the improvement is at most 0.5~dB for all \ac{drr} values. For multiple talkers with a single reflection, the improvement grows from 1.6~dB for a \ac{drr} below 0~dB to 7.2~dB for a \ac{drr} above 10~dB, since with a high \ac{drr} the direct sound of each talker dominates the sound field, while the \ac{lt} selection can upscale only one of the talkers. With three reflections, the improvement is 1.7--2.6~dB for all \ac{drr} values.

\subsubsection{Summary}
Results of the simulation study show that the TF method consistently outperformed the temporal ST and LT method, although the margin in most cases was small, except in multi-talker scenes with few reflections and a high \ac{drr} (Sec.~\ref{subsubsec:multi}). It is also clear from the results that reverberation in term of C50 and IACC are preserved under all conditions, already in FOA, although the additional HOA representation of first reflections does reduce the error. Finally, the LT method although generally showing the worst performance, was superior in reconstructing the direct sound when the number of reflection was at least as high as the number of speakers.

\section{Listening Test}
\label{sec:listening_test}

The simulation results in the previous section show the objective advantage of the \ac{tf} selection, but they do not show whether, and under which conditions, the differences between the selection domains are audible. This section describes a listening experiment designed to compare the perceptual quality of the \ac{tf} approach with that of the temporal methods (\ac{lt} and \ac{st}).

\subsection{Setup}
\label{subsec:lt_setup}
The test signals were generated with the same room simulation, \ac{sh}-domain sound field model and binaural rendering as in the simulation study (Sec.~\ref{subsec:setup}), and the speech signals were taken from the same corpus. Two acoustic scenes from the data set were selected:
\begin{itemize}
    \item \textbf{Scene~1 (single talker):} a female talker at an azimuth of $75^\circ$ and a distance of $0.95$~m in the $18 \times 13.5 \times 5.4$~m room ($T_{60} = 0.8$~s), with a \ac{drr} of $7.5$~dB.
    \item \textbf{Scene~2 (two talkers):} a male and a female talker at azimuths of $\pm 75^\circ$ and a distance of $1.04$~m in the $20 \times 15 \times 6$~m room ($T_{60} = 0.65$~s), with a \ac{drr} of $9.0$~dB.
\end{itemize}

For each scene, five binaural signals were generated: a reference, rendered in full \ac{hoa} of order $N=10$ as in the simulation study; an anchor, rendered entirely in \ac{foa} with the \ac{magls} decoder \cite{schorkhuber2018binaural, zotter2019ambisonics} (equivalent to $r=0$); and the \ac{lt}, \ac{st} and \ac{tf} methods of Sec.~\ref{sec:processing_methods}, with the parameters of Sec.~\ref{subsec:methodology}. The methods were applied with a single upscaled reflection ($r=1$) in both scenes, and with three upscaled reflections ($r=3$) in Scene~2. All signals were loudness-matched to $-26$~LUFS \cite{itu2015loudness}, and their \acs{sdr} relative to the reference, Eq.~\eqref{eq:sdr}, is given in Table~\ref{tab:test_sdr}, which also shows the test screens as detailed in the next section.

\begin{table}[htbp]
    \centering
    \caption{\acs{sdr} (dB) of the listening-test signals relative to the reference.}
    \label{tab:test_sdr}
    \renewcommand{\arraystretch}{1.2}
    \begin{tabular}{@{}lcccc@{}}
    \toprule
    \textbf{Screen} & \textbf{Anchor} & \textbf{\ac{lt}} & \textbf{\ac{st}} & \textbf{\ac{tf}} \\
    \midrule
    1: Single talker, $r=1$ & 4.25 & 16.41 & 16.43 & 16.64 \\
    2: Two talkers, $r=1$ & 2.77 & 4.69 & 10.05 & 15.65 \\
    3: Two talkers, $r=3$ & 2.77 & 16.61 & 12.32 & 19.35 \\
    \bottomrule
    \end{tabular}
\end{table}

\subsection{Methodology}
\label{subsec:lt_methodology}
The study employed the \ac{mushra} test \cite{itu2015mushra}, implemented in MATLAB \cite{matlab2024}. The experiment consisted of three \ac{mushra} screens:
\begin{enumerate}
    \item Single talker, $r=1$, Scene~1.
    \item Two talkers, $r=1$, Scene~2.
    \item Two talkers, $r=3$, Scene~2. 
\end{enumerate}
Each screen included five test signals: the hidden reference, the anchor, and the \ac{lt}, \ac{st} and \ac{tf} signals. Since the \ac{magls} decoder largely removed the timbral differences between the \ac{foa} and \ac{hoa} renderings, the differences between the test signals were mainly spatial. Therefore, the participants were asked to rate the \textit{spatial fidelity} of each test signal relative to the reference, including the source position, width and externalization, on a scale from 0 to 100, where 100 indicates that the test signal is indistinguishable from the reference. Since in the two-talker screens a test signal may affect each talker differently, the participants were instructed to listen to all signals while focusing on one talker, then again while focusing on the other talker, and to give a single rating that accounts for both.

Ten subjects, all reporting normal hearing, participated in the study. All participants are lab members and agreed to take part in the listening test, which was approved by the Human Subjects Research Committee of Ben-Gurion University of the Negev (approval no. 1157-1, August 2, 2025). The experiment was distributed to the participants, and each participant performed it on their own computer and headphones, with the headphone compensation filter of the selected headphone model, in as quiet an environment as possible. The order of the screens and of the test signals within each screen was randomized for every participant. Prior to the test, the participants completed a training stage, in which they learned to use the interface and the rating scale, and became familiar with the full range of test signals.

\subsection{Results}
\label{subsec:lt_results}
Following ITU-R BS.1534-3 \cite{itu2015mushra}, participants who rated the hidden reference below 90 in more than 15\% of the screens were to be excluded. All participants rated the hidden reference 100 in all screens, so that no participant was excluded. A two-way repeated-measures ANOVA with two within-subject factors, method (reference, anchor, \ac{lt}, \ac{st}, \ac{tf}) and acoustic conditions (screens 1, 2 and 3), was performed on the ratings. Mauchly's test indicated a violation of sphericity for both factors; accordingly, the Greenhouse-Geisser correction was applied. The analysis revealed statistically significant main effects of method, $F(1.29, 11.59) = 89.5$, $p < .001$, $\eta_p^2 = .91$, and acoustic conditions, $F(1.24, 11.18) = 20.5$, $p < .001$, $\eta_p^2 = .70$. Furthermore, the interaction between method and screen was significant, $F(2.97, 26.74) = 13.3$, $p < .001$, $\eta_p^2 = .60$, indicating that the differences between the methods depend on the acoustic conditions.

\begin{figure*}[t]
    \centering
    \includegraphics{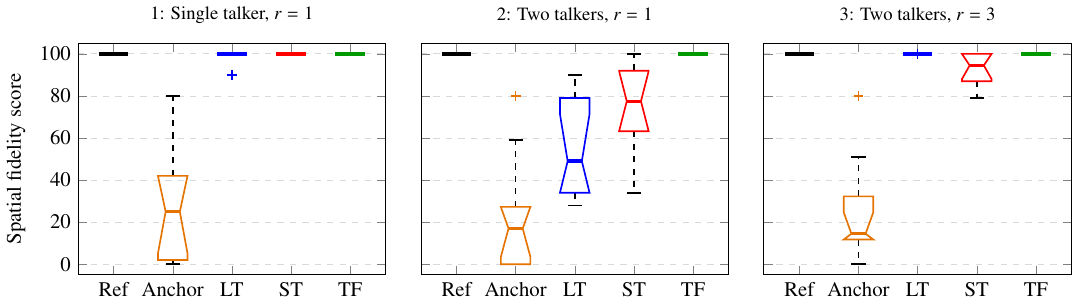}
    \caption{Box plots of the listening-test scores (10 participants) for the hidden reference (Ref), the \ac{foa} anchor, and the \ac{lt}, \ac{st} and \ac{tf} methods, in the three \ac{mushra} screens. Boxes span the 25th--75th percentiles, horizontal lines indicate the median, whiskers extend to the most extreme non-outlying scores, and open circles indicate outliers.}
    \label{fig:listening_test}
\end{figure*}

To further examine the method effect within each screen, post-hoc paired $t$-tests compared every method with the reference, and the three methods with each other (\ac{tf}--\ac{lt}, \ac{tf}--\ac{st}, \ac{st}--\ac{lt}), with a Bonferroni correction within each screen; Wilcoxon signed-rank tests were used to verify the results without assuming normality. The distribution of the participant ratings across methods and screens is presented as box plots in Fig.~\ref{fig:listening_test}. The results can be summarized as follows:
\begin{itemize}
    \item \textbf{Single talker, $r=1$:} all participants rated the \ac{st} and \ac{tf} signals 100, i.e., indistinguishable from the reference, and all but one rated the \ac{lt} signal 100 (mean difference of 1.0 points, $p_{corr} = .69$), while the anchor was clearly identified (mean difference of 73.5 points, $p_{corr} < .001$). This confirms that with a single talker and $r=1$, where all methods presumably select the direct sound, the selection domain has no perceptual effect.
    \item \textbf{Two talkers, $r=1$:} all participants rated the \ac{tf} signal 100, while the \ac{lt} and \ac{st} signals received lower ratings, with mean differences of 44.8 and 24.5 points from the reference ($p_{corr} < .001$ and $p_{corr} = .015$), and the anchor 77.8 points ($p_{corr} < .001$). A direct comparison confirmed that \ac{tf} was rated significantly higher than both \ac{lt} ($p_{corr} < .001$, Wilcoxon $p_{corr} = .006$, Cohen's $d_z = 1.79$) and \ac{st} ($p_{corr} = .015$, Wilcoxon $p_{corr} = .023$, $d_z = 1.17$). The \ac{st} and \ac{lt} methods were not significantly different from each other ($p_{corr} = .052$), and the ratings of both varied considerably between participants. This is consistent with the simulation study, where with a single reflection the temporal methods cannot upscale the direct sound of both talkers, and \ac{lt} upscales the direct sound of only one of them.
    \item \textbf{Two talkers, $r=3$:} all participants rated the \ac{lt} and \ac{tf} signals 100, while the \ac{st} signal was rated slightly lower, with a mean difference of 7.3 points from the reference, at the threshold of significance ($p_{corr} = .050$; the comparison of \ac{tf} and \ac{st} gives $p_{corr} = .050$, but is not confirmed by the Wilcoxon test, $p_{corr} = .094$). With three reflections, \ac{lt} includes the direct sound of both talkers, while the frame-wise selection of \ac{st} may not; this is the order predicted by the \acs{sdr} in Table~\ref{tab:test_sdr}.
\end{itemize}
Overall, the listening-test results corroborate the simulation findings: in a multi-talker scene, a single reflection upscaled in the \ac{tf} domain was perceptually indistinguishable from the \ac{hoa} reference, a quality that the temporal methods reached only with more reflections, while for a single talker all selection domains were equivalent.

\subsection{Discussion}

Note that the two acoustic scenes have a relatively high \ac{drr}. This high DRR was chosen because the simulation study predicted the largest differences between the selection domains for multiple talkers (Fig.~\ref{fig:TF_gain_DRR}) under such conditions. Both single talker and two talkers were included because for a single talker all methods are expected to perform similarly (see Fig.~\ref{fig:Multi_Source_SDR_Global}), while for more talkers discrepancy between the methods became apparent (Figs.~\ref{fig:Multi_Source_SDR_Global} and \ref{fig:TF_gain_DRR}). Furthermore, conditions with a low \ac{drr} were not included in this work, since informal listening tests presented no audible differences between the methods, including the anchor, and the reference signal, which is in agreement with the convergence of the methods for a low \ac{drr} shown in the simulation study (Figs.~\ref{fig:SDR_DRR_Groups} and \ref{fig:TF_gain_DRR}).

\section{Conclusion}
\label{sec:conclusion}

This paper investigated the importance of early reflections in the context of FOA upscaling, addressing a critical gap in current parametric synthesis methods that primarily focus on the direct sound. An analytical framework was proposed to explicitly identify and selectively render prominent early reflections with HOA resolution, leaving the late reverberant tail in FOA. The reflection selection mechanism was evaluated across three distinct processing domains: long-term global energy (LT), short-time frames (ST), and the joint time-frequency (TF) domain.

Comprehensive objective evaluations, utilizing SI-SDR, $C_{50}$, and IACC metrics across a wide range of simulated reverberant environments, demonstrated the distinct advantages of TF-domain processing. By exploiting the inherent sparsity of speech signals, the TF method successfully isolated energetic early reflections without suffering from the temporal masking that degraded the ST approach, or the static bias toward the direct sound observed in the LT approach. This advantage was particularly pronounced in complex, multi-talker scenarios with a high DRR, where the TF method consistently achieved superior reconstruction accuracy and spatial preservation. The listening test confirmed these findings: in a two-talker scene, a single reflection upscaled in TF was indistinguishable from the HOA reference, while the temporal methods required more reflections.

These findings provide critical insights into the spatial upscaling problem, demonstrating that the explicit inclusion and targeted enhancement of early reflections significantly improve the physical accuracy of the reconstructed sound field. The proposed framework highlights the substantial potential of integrating TF-based early reflection reconstruction into the design of future parametric and data-driven FOA upscaling algorithms.

\section*{CRediT authorship contribution statement}
\textbf{Oded Levi:} Performed the research, including the simulation and the listening test, and wrote the paper. \textbf{Boaz Rafaely:} Supervised the research and the writing of the paper.

\section*{Declaration of competing interest}
The authors declare that they have no known competing financial interests or personal relationships that could have appeared to influence the work reported in this paper.

\section*{Funding}
This research did not receive any specific grant from funding agencies in the public, commercial, or not-for-profit sectors.

\section*{Data availability}
The test signals of the listening test and the ratings of the participants (anonymized) are openly available in Zenodo at \url{https://doi.org/10.5281/zenodo.24257365}. The simulation code is available from the corresponding author upon reasonable request.

\section*{Declaration of generative AI and AI-assisted technologies in the manuscript preparation process}
During the preparation of this work the authors used Claude (Anthropic) to assist with language editing. After using this tool, the authors reviewed and edited the content as needed and take full responsibility for the content of the published article.

\bibliographystyle{elsarticle-num}
\bibliography{main}

\end{document}